\documentclass[11pt]{article}
\usepackage{epsf}
\usepackage{citesort}
\usepackage{xspace}

\title{Compositional Competitiveness for Distributed
Algorithms\footnote{An earlier version of this work appeared as 
``Modular competitiveness for distributed algorithms,''
in {\em Proceedings of the Twenty-Eighth Annual ACM Symposium on the
Theory of Computing}, pages 237--246, Philadelphia, Pennsylvania, 22--24 May
1996.}}

\author{James Aspnes\thanks{Yale University, Department~of
Computer Science, 51 Prospect Street/P.O. Box 208285, New
Haven CT 06520-8285. Supported by NSF grants CCR-9410228,
CCR-9415410, CCR-9820888, and CCR-0098078.
E-mail: {\tt aspnes-james@cs.yale.edu}}
\and Orli Waarts\thanks{Computer Science Division,
U. C. Berkeley. Supported in part by an NSF postdoctoral
fellowship. E-Mail: {\tt waarts@cs.berkeley.edu}} }

\newcommand{\done}{\mathop{\rm done}\nolimits}
\newcommand{\opt}{\mathop{\rm opt}\nolimits}
\newcommand{\work}{\mathop{\rm work}\nolimits}

\newtheorem{theorem}{Theorem}
\newtheorem{lemma}[theorem]{Lemma}
\newtheorem{definition}[theorem]{Definition}

\newcommand{\qedsymb}{\hfill{\rule{2mm}{2mm}}}
\newenvironment{proof}{\begin{trivlist}
 \item[\hspace{\labelsep}{\bf\noindent Proof: }]
}{\qedsymb\end{trivlist}}

\newcommand{\buzz}[1]{{\em #1}}

\newcommand{\etal}{{\em et al.}\xspace}

\newcommand{\newloglike}[2]{\newcommand{#1}{\mathop{\rm #2}\nolimits}}
\newloglike{\CL}{CL}
\newloglike{\PL}{PL}

\begin{document}

\maketitle

\begin{abstract}
We define a measure of competitive performance for distributed
algorithms based on \buzz{throughput}, the number of tasks that an
algorithm can carry out in a fixed amount of work. This new
measure complements the \buzz{latency} measure of 
Ajtai~\etal~\cite{AADW-94}, which measures how quickly an algorithm can finish
tasks that start at specified times.  The novel feature of the
throughput measure, which distinguishes it from the latency
measure, is that it is compositional: it supports a notion
of algorithms that are competitive \buzz{relative to} a class of subroutines,
with the property that an algorithm that is $k$-competitive relative
to a class of subroutines, combined with an $\ell$-competitive member of
that class, gives a combined algorithm that is $k\ell$-competitive.

In particular, we prove the throughput-competitiveness of a class of
algorithms for \buzz{collect operations}, in which each of a group of
$n$ processes obtains all values stored in an array of $n$ registers.
Collects are a fundamental building block of a wide variety of
shared-memory distributed algorithms, and we show that several such
algorithms are competitive relative to collects.  Inserting a 
competitive collect in these algorithms gives the first examples
of competitive distributed algorithms obtained by composition using
a general construction.
\end{abstract}

\section{Introduction}
\label{section-introduction}

The tool of competitive analysis was proposed by Sleator and
Tarjan~\cite{ST} to study problems that arise in an \buzz{on-line}
setting, where an algorithm is given an unpredictable sequence of
requests to perform operations, and must make decisions about how to
satisfy its current request that may affect how efficiently it can
satisfy future requests. Since the worst-case performance of an
on-line
algorithm might depend only on very unusual or artificial sequences of
requests, or might even be unbounded if one allows arbitrary request
sequences, one would like to look instead at how well the algorithm
performs relative to some measure of difficulty for the request
sequence. The key innovation of Sleator and Tarjan was to use as a
measure of difficulty the performance of an optimal \buzz{off-line}
algorithm, one allowed to see the entire request sequence before
making any decisions about how to satisfy it. They defined the
\buzz{competitive ratio}, which is the supremum, over all possible input
sequences $\sigma$, of the ratio of the performance achieved by the
on-line algorithm on $\sigma$ to the performance achieved by the optimal
off-line algorithm on $\sigma$, where the measure of performance depends on
the particular problem.

We would like to apply competitive analysis to the design of
asynchronous distributed algorithms, where the input sequence may
reflect both user commands and the timing of events in the underlying
system.  Our goal, following previous work of
Ajtai~\etal~\cite{AADW-94}, is to find
competitive algorithms for core problems in distributed computing that
can then be used to speed up algorithms for solving more complex
problems.  To this end, we need a definition of competitive
performance that permits \buzz{composition}: the construction of a
competitive algorithm by combining a competitive superstructure with a
competitive subroutine.  Our efforts to find such a definition have
led us to a notion of \buzz{competitive throughput}, which counts the
number of operations or tasks that can be completed by some algorithm
in a fixed time frame.  We show that for a particular problem, the
\buzz{cooperative collect problem}, it is possible both (a) to obtain
cooperative collect
algorithms with good competitive throughput and (b) to use these
algorithms as subroutines in many standard algorithms not specifically
designed for competitive analysis, and thereby obtain competitive
versions of these algorithms.

We begin with a short history of competitive analysis in distributed
algorithms (Section~\ref{section-intro-history}), followed by a
discussion of the cooperative collect primitive
(Section~\ref{section-intro-collect}) and an overview of our approach
and the organization of the rest of the paper
(Section~\ref{section-intro-our-approach}).

\subsection{Competitive analysis and distributed algorithms}
\label{section-intro-history}

In a distributed setting, there are additional sources of
nondeterminism other than the request sequence. These include
process step times, request arrival times, message delivery times (in
a message-passing system) and failures. Moreover, a distributed algorithm
has to deal not only with the problems of lack of knowledge
of future requests and future system behavior,
but also with incomplete information about the \emph{current} system
state. Due to the additional type of nondeterminism in the distributed
setting, it is not obvious how to extend the notion of competitive analysis
to this environment.

Early work on distributed job scheduling and data management
\cite{AKP-92,BFR-92,AKRS,AA-94,ABF-93,BR-92,AP-90} took the approach
of comparing a distributed on-line algorithm to a global-control
off-line algorithm. 
However, as noted in \cite{ABF-93}, using a global-control algorithm as a
reference has
the unfortunate side-effect of forcing the on-line algorithm to
compete not only against algorithms that can predict the future but
against algorithms in which each process can deduce what all other
processes are doing at no cost.  

While such a measure can be useful
for algorithms that are primarily concerned with managing resources,
it unfairly penalizes algorithms whose main purpose is propagating
information.
Ajtai~\etal~\cite{AADW-94} described a more refined approach in which
a \buzz{candidate} distributed algorithm is compared to an optimal
\buzz{champion}.  In their \buzz{competitive latency}
model, both the candidate and champion are
distributed algorithms.  Both are subject to an unpredictable
\buzz{schedule} of events in the system and both must satisfy the same
correctness condition for all possible schedules.  The difference is
that the adversary may supply a different champion optimized for each 
individual schedule when measuring performance.

\subsection{Cooperative collect}
\label{section-intro-collect}

The competitive latency model was initially
designed to analyze a particular
problem in distributed computing called \buzz{cooperative collect},
first abstracted by Saks~\etal~\cite{SSW}.  The cooperative collect
problem arises in asynchronous shared-memory systems built from
single-writer registers.\footnote{A single-writer register is one that
is ``owned'' by some process and can only be written to by its
owner.}  In order to observe the state of the system, a process must
read $n-1$ registers, one for each of the other processes.  
The simplest
implementation of this operation is to have the process carry out
these $n-1$ reads by itself.  
However, if many processes are trying
to read the same set of registers simultaneously, some of
the work may be shared between them.

A \buzz{collect operation} is any procedure by which a process may
obtain the values of a set of $n$ registers (including its own).  The
correctness conditions for a collect are those that follow naturally
from the trivial implementation consisting of $n-1$ reads.  The process
must obtain values for each register, and those values must be
\buzz{fresh}, meaning that they were
present in the register at some time between when the collect started
and the collect finished.

Curiously, the trivial implementation is the one used in almost all of
the many asynchronous shared-memory algorithms based on collects,
including algorithms for consensus, snapshots, coin flipping, bounded
round numbers, timestamps, and multi-writer
registers~\cite{Ab,G6,And,Asp,AH,%
AHSnap,AspW,AHR,AR,BG-93,BrR,CIL,DolSh,DHPW,DHW,DW,
GLS,Haldar93,H,IL,IsPin,KirousisST1996,VA}.  (Noteworthy exceptions are 
\cite{SSW,RST}, which present
interesting collect algorithms that do not follow the pattern of the
trivial algorithm, but which depend on making strong
assumptions about the schedule.)
Part of the reason for the popularity of this approach may be that the
trivial algorithm is optimal in the worst case: a process running in
isolation has no alternative but to gather all $n-1$ register values
itself.

Ajtai \etal's~\cite{AADW-94} 
hope was that a cooperative collect subroutine with a
good competitive ratio would make any algorithm that used it run
faster, at least in situations where the competitive ratio implies
that the subroutine outperforms the trivial algorithm.  
To this end, they constructed the first known competitive algorithm for
cooperative collect and showed a bound on its competitive latency.
Unfortunately,
there are technical obstacles in the competitive latency model that
make it impossible to \emph{prove} that an algorithm that uses a
competitive collect is itself competitive.  The main problem is that
the competitive latency includes too much information in the schedule:
in addition to controlling the timing of events in the underlying
system such as when register operations complete, it specifies when
high-level operations such as collects begin.  So the competitive
latency model can only compare a high-level algorithm to
other high-level algorithms that use collects at exactly the same
times and in exactly the same ways.

\subsection{Our approach}
\label{section-intro-our-approach}

In the present work, we address this difficulty by replacing the
competitive latency measure with a \buzz{competitive throughput}
measure that assumes that the candidate and champion face the same
behavior in the system, but breaks the connection between the tasks
carried out by the candidate and champion algorithms.
This model is described in detail in
Section~\ref{section-competitive-throughput}.
The intuition
is that when analyzing a distributed algorithm it may be helpful to
distinguish between two sources of nondeterminism, user requests (the
input) and system behavior (the schedule).  Previous work that compares a
distributed algorithm with a global control algorithm
\cite{AKRS,AA-94,ABF-93,AKP-92,AP-90,BFR-92,BR-92} implicitly makes this
distinction by having the on-line and off-line algorithms compete only on
the same input, generally hiding the details of the schedule in a
worst-case assumption applied only to the on-line algorithm. 
In effect, these models use a competitive input but a worst-case
schedule.
The
competitive latency model of~\cite{AADW-94} applies the same input and
schedule to both the on-line and the off-line algorithms. 
In contrast, we assume a worst-case {\em input} but a competitive {\em
schedule}.  
Assuming a worst-case input means an algorithm must perform well in
any context--- including as part of a larger algorithm.  At the same
time, comparing the algorithm to others with the same schedule gives
a more refined measure of the algorithm's response to bad system
behavior than a pure worst-case approach.

The competitive throughput model 
solves
the problem of comparing an algorithm $A$ using a
competitive subroutine $B$ against an algorithm $A^*$ that uses a
subroutine $B^*$ implementing the same underlying task, but it does not
say anything about what happens when comparing $A$ to an optimal $A^*$ that
does not call $B^*$.  For this we need an additional tool that we call
\buzz{relative competitiveness}, described in
Section~\ref{section-composition-of-competitive-algorithms}.
We show (Theorem~\ref{theorem-composition}) that if an algorithm $A$ is
$k$-relative-competitive with respect to an underlying subroutine $B$, 
and $B$ is itself $l$-competitive, then the combined algorithm $A\circ
B$ is $kl$-competitive, even against optimal algorithms $A^*$ that do not use
$B$.

To demonstrate the applicability of these techniques, we show in
Section~\ref{section-cooperative-collects} that the results of
\cite{AADW-94} can be extended to bound the competitive throughput of
their algorithm; in fact, our techniques apply to any algorithm for
which we have a bound on an underlying quantity that \cite{AADW-94}
called the \buzz{collective latency}, a measure of the total work
needed to finish all tasks in progress at any given time.
(This result has been used since
the conference appearance of the present
work by Aspnes and Hurwood~\cite{AspnesH1998}
to prove low competitive throughput for an algorithm that improves on
the algorithm of \cite{AADW-94}.)
We show in Section~\ref{section-applications}
that relative competitiveness, combined with a throughput-competitive
collect algorithm, does in fact give throughput-competitive
solutions to problems such as atomic snapshot~\cite{G6,And,AHSnap,AHR,AR}
and
bounded round numbers \cite{DHW}.  We argue that most
algorithms that use collects can be shown to be throughput-competitive
using similar techniques.

Finally, in Section~\ref{section-conclusions} we discuss some related
approaches to analyzing distributed algorithms and consider what
questions remain open.

\section{Model}
\label{section-model}

We use as our underlying model the wait-free shared-memory model of
\cite{Herlihy91}, using atomic single-writer multi-reader registers as
the means of communication between processes.
Because the registers are atomic, we can represent an execution as an
interleaved sequence of steps, each of which is a read or write of
some register.  The timing of events in the system is
assumed to be under the control of an adversary, who is allowed to see
the entire state of the system (including the internal states of the
processes
and the contents of the registers).
The adversary decides at each time unit which process
gets to take the next step; these decisions are summarized in 
a \buzz{schedule}, 
which formally is just a sequence of process id's.  

The algorithms we consider implement
\buzz{objects}, which are abstract concurrent data structures with
well-defined interfaces and correctness conditions.
We assume that:
\begin{enumerate}
\item The objects are manipulated by invoking \buzz{tasks} of some
sort; 
\item That each instance of a task has a well-defined \buzz{initial
operation} and a well-defined \buzz{final operation} (which may equal
the initial operation for simple tasks).
\item That the definition of an initial or final operation depends
only on the operation and the preceding parts of the execution, so
that the completion of a task is recognizable at the particular step of the
execution in which its final operation is executed, without needing to
observe any subsequent part of the execution, and so that the number
of tasks completed in an execution can be defined simply by counting
the number of final operations executed; and
\item That there is a predicate on object schedules that distinguishes
correct
executions from incorrect executions,
so that correct implementations are defined as those whose executions
always satisfy this correctness predicate.  
\end{enumerate}
Beyond these minimal
assumptions, the details
of objects will be left unspecified unless we are dealing with
specific applications.  

Each process has as input a \buzz{request sequence} specifying what
tasks it must carry out.  The request sequences are supplied by the
adversary and are \emph{not} part of the schedule, as we may wish to
consider the effect of different request sequences while keeping the
same schedule.  We assume that the request sequences are long enough
that a process never runs out of tasks to perform.

The performance of an algorithm is measured by its competitive
throughput, defined in Section~\ref{section-competitive-throughput}.
We contrast this definition with the competitive latency measure of
Ajtai~\etal{}~\cite{AADW-94} in
Section~\ref{section-comparison-with-competitive-latency}.
Building throughput-competitive algorithms by
composition is described in
Section~\ref{section-composition-of-competitive-algorithms}.

\section{Competitive throughput}
\label{section-competitive-throughput}

\begin{figure*}[t]
\epsfbox{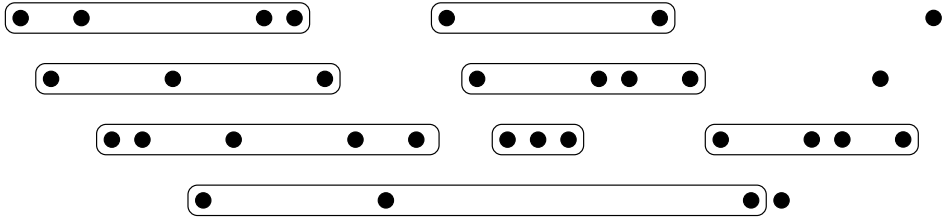}
\caption{\label{figure-throughput}
Throughput model.  High-level operations (ovals) are implemented as a
sequence of low-level steps (circles), which take place at times
determined by the adversary.  New high-level operations start as
soon as previous operations end.  Payoff to the algorithm is number
of high-level operations completed.}
\end{figure*}

The competitive throughput of an algorithm measures how many tasks an
algorithm can complete with a given schedule.  The assumption is that
each process starts a new task as soon as each previous task is
finished, as shown in Figure~\ref{figure-throughput}.

We measure the algorithm against a champion algorithm that runs under
the same schedule.  We do not assume that both algorithms are given
the same request sequences; we only require that the two sets of
request sequences be made up of tasks for the same object $T$.  This
assumption may seem unfair to the candidate algorithm, but it is
necessary to allow algorithms to be composed.  In reasoning about
competitiveness compositionally, we compare the efficiency of a
candidate $B$ used as a subroutine in some higher-level algorithm $A$
with the champion $B^*$ used as a subroutine in some optimal
higher-level algorithm $A^*$.  In general we do not expect $A$ and
$A^*$ to generate the same request sequences to $B$ and $B^*$ (hence
the split between worst-case request sequences for $B$ and best-case
for $B^*$), but we
can insist that both $B$ and $B^*$ run under the same schedule.

We start by introducing some notation.
For each algorithm $A$, schedule $\sigma$, and set of request
sequences $R$, define $\done(A,\sigma,R)$ to be the total number of
tasks completed by all processes when running $A$ according to the
schedule $\sigma$ and set of request sequences $R$.
Define $\opt_{T}(\sigma)$
to be $\max_{A^*,R^*} \done(A^*,\sigma,R^*)$, where $A^*$
ranges over all correct
implementations of $T$ and $R^*$ ranges over
all sets of request sequences composed of
$T$-tasks.  (Thus, $\opt_{T}(\sigma)$ represents the
performance of the best correct algorithm running on the best-case
request sequences for the fixed schedule $\sigma$.)

\begin{definition}
\label{definition-competitive-throughput}
Let $A$ be an algorithm that implements an object $T$.
Then $A$ is \buzz{$k$-throughput-competitive}
for $T$ if there exists a constant $c$
such that, for any schedule $\sigma$ and set of request sequences $R$,
\begin{equation}
\label{eq-competitiveness}
\done(A,\sigma,R) + c
 \geq {1 \over k} \opt_{T}(\sigma).
\end{equation}
\end{definition}

This definition follows the usual definition of 
competitive ratio~\cite{ST}.
The ratio is inverted since $\done(A,\sigma,R)$ measures a payoff
(the number of completed tasks) to be maximized instead of a cost to
be minimized.
The constant $c$ is included to avoid problems that would otherwise
arise from the granularity of tasks.  On very short schedules, it
might be impossible for $A$ to complete even a single task, even
though the best $A^*$ could.  Allowing the constant (which has minimal
effect on longer schedules) gives a measure that more realistically
describes the performance of $A$ on typical schedules.

\subsection{Comparison with competitive latency}
\label{section-comparison-with-competitive-latency}

The competitive throughput measure was inspired by the similar
competitive latency measure of
Ajtai~\etal~\cite{AADW-94}.
Competitive latency is not used in this paper, but we will give the
definition from~\cite{AADW-94} to permit direct
comparison between the two measures.

\begin{figure*}[t]
\epsfbox{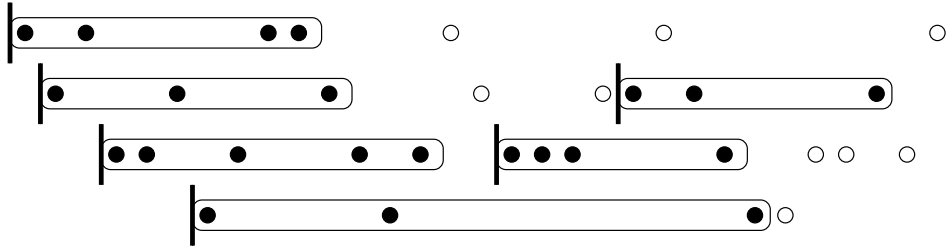}
\caption{\label{figure-latency}
Latency model.  New tasks (ovals) start at
times specified by the schedule (vertical bars).  Schedule also
specifies timing of
low-level steps (small circles).  Cost to algorithm is number of
low-level operations actually performed (filled circles), ignoring
time steps allocated to processes in between tasks (empty circles).}
\end{figure*}

In the competitive latency model, the request sequences, including the
times at which tasks start, are
included in the schedule (see Figure~\ref{figure-latency}).
Thus the schedule $\sigma$ includes both user input (the request
sequences) and system timing (when each process is allowed to take a
step).
It is assumed that each task runs to completion, and that the process
executing the task becomes idle until its next task starts; if the
schedule calls for the process to carry out a step in between tasks,
it performs a noop.
The \buzz{total work} done by an algorithm $A$ given
a schedule $\sigma$, written $\work(A,\sigma)$,
is defined as the number of operations performed
by processes outside of their idle periods.  

The \buzz{competitive latency} of a candidate algorithm $A$ is defined
as~\cite{AADW-94}:
\begin{displaymath}
\sup_{\sigma} {{\work(A,\sigma) \over \inf_{A^*} \work(A^*,\sigma)}},
\end{displaymath}
where $\sigma$ ranges over all schedules in which every task demanded
of $A$ is given enough steps to finish before a new task starts.
We can rewrite this in a form closer to that of
Definition~\ref{definition-competitive-throughput}, by writing that an
algorithm $A$ is $k$-latency-competitive if, for all schedules
$\sigma$ that permit $A$ to finish its tasks, and all champion
algorithms $A^*$,
\begin{displaymath}
\work(A,\sigma) \le k \work(A^*,\sigma).
\end{displaymath}

Note that this definition does not include an additive constant.  
The definition of competitive throughput does, to avoid problems with
very short schedules in which the candidate $A$ cannot complete any
tasks.  This problem does not arise with the competitive latency
definition because of the restriction to schedules in which $A$
can complete all assigned tasks.

Competitive latency has some advantages over competitive throughput.
Because the request sequences are part of the schedule, it can be used
to evaluate algorithms for which some tasks are much more expensive
than others.  
In such a situation,
the candidate in the competitive throughput model 
may be stuck
with hard tasks while the champion breezes through easy ones.
Competitive latency may thus be a better model than competitive
throughput for measuring the
performance of algorithms in isolation, though competitive throughput
is a better measure for subroutines, as it allows composition
using the results in
Section~\ref{section-composition-of-competitive-algorithms}.
Often, difficulties with varying costs can also
be ameliorated by joining cheap tasks to
subsequent expensive ones, as is done in Sections~\ref{section-write-collect}
and \ref{section-snapshot}.

On the other hand, competitive throughput removes some awkward
features of competitive latency.  In particular, the assumption that
processes are idle in between tasks (which is a necessary side-effect
of specifying when in the sequence of operations each new task starts)
may not be a good representation of how distributed algorithms are
implemented in practice.  A further concern is that if a process
becomes idle quickly (say, during a brief lenient period in the
schedule that allows quick termination), it is unaffected by harsher
conditions that may arise later.  This means in particular that a
candidate algorithm that is only slightly slower than the champion it
is competing against may find itself operating under much worse
conditions, as the schedule suddenly gets worse as soon as the
champion finishes.

The contrast between competitive latency and competitive throughput
suggests a trade-off between competing notions of fair competition.
Competitive latency treats different algorithms unfairly with respect
to the schedule, by forcing a candidate to continue running in bad
conditions after the champion has finished in good ones.  But
competitive throughput treats different algorithms unfairly with
respect to the request sequences, because the definition explicitly
assumes that the requests given to the candidate and champion may be
different.  It is not clear whether a more sophisticated definition
could avoid both extremes, and produce a more accurate measure of the
performance of a distributed algorithm compared to others running
under similar conditions.

\section{Composition of competitive algorithms}
\label{section-composition-of-competitive-algorithms}

The full power of the
competitive throughput measure only becomes apparent when we consider
competitive algorithms built from competitive subroutines.
In traditional worst-case analysis, an algorithm that invokes a
subroutine $k$ times at a cost of at most $l$ 
time units each pays a total of $kl$ time units.
In a competitive framework, both the number of times the subroutine is
called and the cost of each call to the subroutine may depend on
system nondeterminism.
The analogous quantity to the cost $l$ of each subroutine call is the
competitive ratio of the subroutine.  What is an appropriate analog of
the number of times $k$ that the subroutine is called?

In Section~\ref{section-relative-competitiveness}, we define a notion
of \buzz{relative competitiveness} that characterizes how well an
algorithm uses a competitive subroutine.
As shown in Section~\ref{section-composition-theorem},
algorithms that are $k$-throughput-competitive relative to an
$l$-throughput-competitive subroutine are themselves
throughput-competitive, with ratio $kl$.
The definition of relative competitiveness
(Definition~\ref{definition-relative-competitiveness})
and the composition theorem that uses it
(Theorem~\ref{theorem-composition})
yield a method for constructing competitive algorithms compositionally.
Some examples of applications of this method appear in
Section~\ref{section-applications}.
To our knowledge this is the first example of a general composition
theorem for competitive algorithms, even outside of a distributed
setting.

\subsection{Relative competitiveness}
\label{section-relative-competitiveness}

As in the definition of throughput-competitiveness, we consider a
situation in which $A$ is an algorithm implementing some object $T$.
Here, however, we assume that $A$ depends on a
(possibly unspecified) subroutine implementing a different object $U$.
For any specific algorithm $B$ that implements $U$, we will
write $A \circ B$ for the composition of $A$ with $B$, i.e., for that
algorithm which is obtained by running $B$ whenever $A$ needs to
carry out a $U$-task.\footnote{For this definition it is important that $A$
not execute any operations that are not provided by
$U$.  In practice, the difficulties
this restriction might cause can often be avoided by treating $U$ as a
composite of several different objects.}

\begin{definition}
\label{definition-relative-competitiveness}
An algorithm $A$ is \buzz{$k$-throughput-competitive for $T$ 
relative to $U$} if there exists a constant $c$ such that for any $B$
that implements $U$, and any schedule $\sigma$ and request sequence
$R$ for which the
ratios are defined,
\begin{equation}
\label{eq-relative-competitiveness}
\frac{\done(A\circ B,\sigma,R) + c}{\done(B,\sigma,R_A)}
\geq \frac{1}{k} 
  \cdot
     \frac{\opt_{T}(\sigma)}
          {\opt_{U}(\sigma)},
\end{equation}
where $R_A$ is the request sequence
corresponding to the subroutine calls in $A$ when running according to
$R$ and $\sigma$.
\end{definition}

As in the preceding definition, the additive constant
$c$ is included to  avoid problems with granularity.  The condition that
the ratios are defined, which in essence is just a requirement that
$\sigma$ be long enough for $B$ to complete at least one $U$-task, is
needed for the same reason.

The condition that the ratios are defined in 
(\ref{eq-relative-competitiveness})
does create a curious loophole in the definition of
relative competitiveness: if $A$ implements some object $T$ using
an object $U$ whose tasks can never be completed by a correct
implementation, then the denominators in the inequality 
(\ref{eq-relative-competitiveness}) 
are always zero, and 
thus $A$ is vacuously
zero-competitive relative to $U$.  
Similarly, an implementation $B$ of $U$ that never completes any tasks
will be vacuously zero-competitive for $U$.  Since $A \circ B$ is
unlikely to be zero-competitive for $T$, 
to apply relative competitiveness
we will need to exclude such pathologies.  We do so using the
following definition of \buzz{relative feasibility} of objects:
\begin{definition}
\label{definition-relative-feasibility}
Let $T$ and $U$ be objects.  Say that $T$ is \buzz{feasible
relative to} $U$ 
if there exists a constant $c$ such that 
for all schedules $\sigma$,
\begin{equation}
\label{eq-relative-feasibility}
\opt_{T}(\sigma) \leq c \cdot \opt_{U}(\sigma)
\end{equation}
\end{definition}
In particular, if Definition~\ref{definition-relative-feasibility}
holds, then in any schedule where $A$ completes at least one
operation, $\opt_{U}$ is not zero and the right-hand side of 
(\ref{eq-relative-competitiveness}) is defined.  Furthermore, if $B$
is competitive relative to $U$, then $\done(B,\sigma,R_A)$ is also
nonzero for sufficiently long schedules, and the left-hand side of
(\ref{eq-relative-competitiveness}) is
also defined.

\subsection{The composition theorem}
\label{section-composition-theorem}

Theorem~\ref{theorem-composition} describes under what conditions a
relative-competitive algorithm combines with a competitive subroutine
to yield a competitive algorithm.

\begin{theorem}
\label{theorem-composition}
Let $A$ be an algorithm that is $k$-throughput-competitive for $T$
relative to $U$, where $T$ is feasible relative to $U$.
Let $B$ be an $l$-throughput-competitive algorithm
for $U$.
Then $A \circ B$ is $kl$-throughput-competitive for $T$.
\end{theorem}
\begin{proof}
For the most part the proof requires only very simple algebraic
manipulation of the definitions, but we must be careful about the
constants and avoiding division by zero.

Fix $\sigma$ and $R$.
We can rewrite the inequality (\ref{eq-relative-competitiveness})
as
\begin{equation}
\label{eqn: theorem composition rel comp}
\left(\done(A,\sigma,R) + c_A\right)\opt_U(\sigma)
 \geq \frac{1}{k} \opt_T(\sigma) \done(B,\sigma,R_A),
\end{equation}
where $c_A$ is the constant from the definition of relative
competitiveness for $A$.  Note that
(\ref{eqn: theorem composition rel comp})
holds even if one or both of the ratios in
(\ref{eq-relative-competitiveness})
is undefined, since in that case $\done(B,\sigma,R_A)$ 
must be zero and all other quantities are non-negative.

We can similarly rewrite (\ref{eq-competitiveness}) as
\begin{equation}
\label{eqn: theorem composition comp}
\done(B,\sigma,R_A) \geq \frac{1}{l}\opt_U(\sigma) - c_B
\end{equation}
where $c_B$ is a constant independent of $\sigma$ and
$R$.  Plugging 
(\ref{eqn: theorem composition comp})
into the right-hand side of 
(\ref{eqn: theorem composition rel comp}) gives
\begin{displaymath}
\left(\done(A,\sigma,R) + c_A\right)\opt_U(\sigma)
 \geq \frac{1}{kl} \opt_T(\sigma) \opt_U(\sigma)
      - \frac{c_B}{k} \opt_T(\sigma),
\end{displaymath}
which combines with the relative feasibility condition
$\opt_T(\sigma) \leq c_T \opt_U(\sigma)$ to give
\begin{displaymath}
\left(\done(A,\sigma,R) + c_A\right)\opt_U(\sigma)
 \geq \frac{1}{kl} \opt_T(\sigma) \opt_U(\sigma)
      - \frac{c_B c_T}{k} \opt_U(\sigma).
\end{displaymath}

This last inequality gives the desired result, as either 
$\opt_U(\sigma) > 0$ and we can divide out $\opt_U(\sigma)$,
or $\opt_U(\sigma) = 0$ and thus $\opt_T(\sigma) = 0$.  In
either case we have
\begin{equation}
\label{eq-composition-final}
\done(A,\sigma,R) + c_A + \frac{c_B c_T}{k} \geq \frac{1}{kl} \opt_T(\sigma).
\end{equation}
\end{proof}

Since we have dropped no terms in this derivation, if
each of the inequalities 
(\ref{eq-competitiveness}),
(\ref{eq-relative-competitiveness}),
and
(\ref{eq-relative-feasibility}) used in the proof is tight,
then (\ref{eq-composition-final}) is also tight;
so the additive constant
$c_A + \frac{c_B c_T}{k}$ is the best possible that can be obtained
without using additional information.


\section{Cooperative collects}
\label{section-cooperative-collects}

In this section, we define the \buzz{write-collect object}, which
encapsulates the cooperative collect problem, and show how any
cooperative collect algorithm satisfying certain natural criteria is
throughput-competitive.

\subsection{The write-collect object}
\label{section-write-collect}

The write-collect object acts like a set of $n$ single-writer
$n$-reader atomic registers
and provides two operations for manipulating these registers.

\begin{enumerate}
\item A
\buzz{collect} operation returns the values of all of the registers,
with the guarantee that any value returned was not overwritten before
the start of the collect. 
\item A \buzz{write-collect} operation writes a new
value to the process's register and then performs a collect.  
\end{enumerate}

The write-collect operation
must satisfy a rather weak serialization condition.  Given two
write-collects $a$ and $b$:
\begin{itemize}
\item If the first operation of $a$ precedes the first operation of
$b$, then $b$ returns the value written by $a$ as part of its vector.
\item If the first operation of $a$ follows the last operation of $b$,
then $b$ does not return the value written by $a$.
\item If the first operation of $a$ occurs during the execution of
$b$, $b$ may return either the value written by $a$, or the previous
value in the register written to by $a$.
\end{itemize}
A trivial implementation of write-collect
might consist of a write followed immediately
by $n$ reads.

Our definition of the write-collect operation is
motivated by the fact that many shared-memory algorithms execute collects
interspersed with write operations (some examples are given in Section
\ref{section-applications}).  Treating write and collect as separate
operations, though in many ways a more natural approach,
also leads to difficulties in applying competitive throughput, as a
candidate doing only expensive collects might find itself in competition with a
champion doing only cheap writes.

\subsection{Competitive algorithms for write-collect}
\label{section-implementation-of-write-collect}

To implement a write-collect, we start with the cooperative collect
algorithm of
\cite{AADW-94}.  This algorithm has several desirable properties,
shown in \cite{AADW-94}:
\begin{enumerate}
\item All communication is through a set of single-writer registers,
one for each process, and the first step of each collect operation is a
write.

\item No collect operation ever requires more than $2n$ steps to
complete.

\item For any schedule, and any set of collects that are in progress
at some time $t$, there is a bound of $O(n^{3/2} \log^2 n)$
on the total number of steps
required to complete these collects.  
\end{enumerate}

These properties are what we need from a cooperative collect
implementation to prove that it gives a throughput-competitive
write-collect.  The first property allows us to ignore the
distinction between collect and write-collect operations (at least in
the candidate): we can include the value written by
the write-collect along with this initial write, and thus trivially
extend a collect to a write-collect with no change in the behavior
of the algorithm.  In effect, our throughput-competitive write-collect
algorithm is simply the latency-competitive collect of \cite{AADW-94},
augmented by merging the write in a write-collect with the first write
done as part of the collect implementation.

The last two properties give two complementary bounds on the number of
steps needed to finish collects in progress at any given time.
The bound on the total work to finish a set of simultaneous collects,
called the \buzz{collective latency}, shows that processes can combine
their efforts effectively when many are running simultaneously.  The
bound on the work done by any individual process, called the
\buzz{private latency}, applies when
only a few processes are running.
We show, in Section~\ref{section-competitive-proof},
that any algorithm $A$ 
with collective latency $\CL(A)$ and private latency $O(n)$ is
$O\left(\sqrt{CL(A)}\right)$-throughput-competitive.

\begin{sloppypar}
For the algorithm of \cite{AADW-94}, the proof gives a competitive
ratio of $O(n^{3/4} \log n)$.  This is the best algorithm currently
known for doing collects in a model in which the adversary has
complete knowledge of the current state of the system.  
It is likely that better algorithms are possible, even in this strong
model,
although the analysis of cooperative collect algorithms can be very
difficult.
\end{sloppypar}

Other authors have devised faster algorithms for weaker models;
see Section~\ref{section-improvements}.

\subsection{Proving throughput-competitiveness}
\label{section-competitive-proof}

We measure time by the total number of steps taken by all
processes.  Consider some execution, and let $C(t)$ be the set of
collect operations in progress at any time $t$.  
Each collect operation consists of a sequence of atomic read and write
operations; if, for some algorithm $A$,
there is a bound $\CL(A)$ on the total number of read and
write operations performed by collects in $C(t)$ at time $t$ or later,
this bound is called the \buzz{collective latency} of
$A$~\cite{AADW-94}.

Define the \buzz{private latency} $\PL(A)$ of $A$ as the maximum
number of read and write operations carried out by a single process
during any one of its own collect operations.

Our progress measure keeps track of how much of the collective latency
and private latency is used up by each read or write operation.  
It is composed of two parts for each process $p$: the first part,
$M_p$, tracks steps by other processes that contribute
to the collective latency of a set of collects that includes $p$'s
current collect.  The second part, $N_p$, simply counts the
number of steps done by $p$.

A step $\pi$ at time $t$
by a process $q$ is \buzz{useful} for $p$ 
if
$\pi$ is part of a collect that started before $p$'s current
collect.  In a sense, $\pi$ is useful if it
contributes to the total work done
by all collects in progress when $p$'s current collect started.
A step $\pi$ is \buzz{extraneous} for $p$ if
$\pi$ occurs during an interval where $p$ has finished one
collect operation but has not yet taken any steps as part of a new
collect operation, and
$\pi$ is either the first or last operation of $q$ in this interval.
Extraneous steps do not help $p$ in any way, but we must count them
anyway for technical reasons that will become apparent in the proof of
Lemma~\ref{lemma-progress-rises}.

Let $M_p(t)$ be the total number of useful and extraneous
steps for $p$ in the first $t$
steps of the execution.  Let $N_p(t)$ be the number of steps carried
out by $p$ in the first $t$ steps of the execution.

\begin{lemma}
\label{lemma-progress-implies-collects}
Let $A$ be any cooperative collect algorithm for which $\CL(A)$ and
$\PL(A)$ are bounded.  Then, in any execution of $A$ in which process
$p$ completes a collect at time $t$, the total number of collects
completed by $p$ by time $t$ is at least $F_p(t) - 1$, where
\begin{equation}
F_p(t) =
\label{eq-progress}
\frac{1}{2}
\left(
    \frac{M_p(t)}{\CL(A) + 2(n-1)}
    +
    \frac{N_p(t)}{\PL(A)}
\right)
\end{equation}
\end{lemma}
\begin{proof}
Observe that $F_p$ does not decrease over time.
We will show that $F_p(t)$ rises by at most $1$ during any single
collect operation of $p$, from which the stated bound will follow.

Define $t_0 = 0$, so that $F_p(t_0) = M_p(t_0) = N_p(t_0) = 0$, 
and for each $i > 0$
let $t_i$ be the time of the last step
of $p$'s $i$-th collect.  We will bound the increase in $N_p$ and
$M_p$ between $t_i$ and $t_{i+1}$ separately.

Since $p$ performs at most $\PL(A)$ steps during 
each collect, we have $N_p(t_{i+1}) - N_p(t_i) \le \PL(A)$. 

Recall that $M_p$ counts both \emph{useful} steps for $p$ (those
operations of other processes $q$ that occur during a collect of $p$
and are part of a collect of $q$ that started before the collect of
$p$) and \emph{extraneous} steps for $p$ (the first and last steps of
any other process $q$ between two successive collects of $p$).
The interval $(t_i,t_{i+1}]$ includes both an initial prefix before
$t$ starts its $(i+1)$-th collect and a suffix during which it carries
out its $(i+1)$-th collect.  
Let $s_{i+1}$ be the time of the first step of $p$'s $(i+1)$-th
collect.
Then during the initial prefix $(t_{i},s_{i+1})$
each other process $q$ may carry out up to two extraneous
steps, for a total of at most $2(n-1)$ extraneous steps.  
Useful steps occur only during the suffix $[s_{i+1},t_{i+1}]$,
and any useful step done by some process $q \ne p$, 
by definition, is part of a
collect that has already started at time $s_{i+1}$.
Since collects in progress
at time $s_{i+1}$ perform a total of at most $\CL(A)$ steps after time
$s_{i+1}$, the total number of useful steps in the interval
$(t_{i},t_{i+1}]$ is at most $\CL(A)$.  Adding together the
useful steps and the extraneous steps gives
$M_p(t_{i+1} - M_p(t_i) \le \CL(A) + 2(n-1)$.

Thus
\begin{eqnarray*}
F_p(t_{i+1}) - F_p(t_i)
&\le&
\frac{1}{2}
\left(
\frac{M_p(t_{i+1}) - M_p(t_i)}{\CL(A) + 2(n-1)}
+
\frac{N_p(t_{i+1}) - N_p(t_i)}{\PL(A)}
\right)
\\
&\le&
\frac{1}{2}\left(1 + 1\right)
= 1.
\end{eqnarray*}

Starting with $F_p(t_0) = 0$, a simple induction argument then shows
that $F_p(t_i) \le i$ for all $i$.

We now exploit the fact that $F_p$ is nondecreasing over time (which
is immediate from the definitions of $M_p$ and $N_p$ and the fact that
$F_p$ is an increasing function of $M_p$ and $N_p$).
Fix some time $t$.  Let $i$ be the largest integer for which $F_p(t) >
i$, so that we have $F_p(t) \le i+1$ or $i \ge F_p(t) - 1$.
If $t < t_i$, then $F_p(t) \le F_p(t_i) \le i$.  Taking the
contrapositive, if $F_p(t) > i$, then $t \ge t_i$.  Since $t_i$ is
defined as the completion time of $p$'s $i$-th collect, $p$ completes
at least $i \ge F_p(t) - 1$ collects by time $t$.
\end{proof}

Now we must show that our progress measure rises.  It is easy to see
that $\sum_p N_p$ rises by exactly 1 per step.  To show that $\sum_p M_p$
rises, we partition the schedule into intervals
and look at how many processors are
active during each interval.

\begin{lemma}
\label{lemma-progress-rises}
Fix some collect algorithm $A$.
Let $(t_1,t_2]$ be any time interval,
and suppose that there are exactly $m$ processes that carry out at least one
step during $(t_1,t_2]$.
Then
\begin{equation}
\label{eq-progress-rises}
\sum_p M_p(t_2) - \sum_p M_p(t_1)
\ge
{m \choose 2}.
\end{equation}
\end{lemma}
\begin{proof}
Recall that $M_p$ counts the total number of steps that are useful for
$p$ or extraneous for $p$.  It is easy to see that, for any $p$,
$M_p(t_2) - M_p(t_1) \ge 0$; this will allow us to ignore processes
that do no take steps during the interval.
For every \emph{pair} of processes that both take steps
during
the interval, we will show that
at least one step of one of the processes is either
useful or extraneous for the other, and thus raises $M_p$ by $1$ for
some $p$.  This will give the desired bound by counting the number of
such pairs.

Let $S$ be the set of processes that carry out at least one step in
$(t_1,t_2]$.
Given distinct processes $p_1,p_2 \in S$,
Define the indicator variable $m_{p_1p_2}$ to equal $1$ if $p_2$ takes at
least one step during $(t_1,t_2]$ that is either useful or extraneous
for $p_1$.  Observe that for any $p_1 \in S$,
\begin{displaymath}
M_{p_1}(t_2) - M_{p_1}(t_1) \ge \sum_{p_2 \in S, p_2 \ne p_1} m_{p_1p_2},
\end{displaymath}
from which it follows that
\begin{eqnarray}
\sum_p M_p(t_2) - \sum_p M_p(t_1)
&\ge& \sum_{p_1 \in S} M_{p_1}(t_2) - \sum_{p_1 \in S} M_{p_1}(t_1) 
\nonumber
\\
&\ge& \sum_{p_1 \in S} \sum_{p_2 \in S, p_2 \ne p_1} m_{p_1p_2} 
\nonumber
\\
&=& \sum_{p_1,p_2 \in S, p_1 \ne p_2} (m_{p_1p_2} + m_{p_2p_1}).
\label{eq-lemma-progress-rises}
\end{eqnarray}

We will now show that, for each distinct pair of processes $p_1,p_2$ 
in $S$, $m_{p_1p_2}+m_{p_2p_1}
\ge 1$.

Let $p_1$ and $p_2$ be processes that take steps in $(t_1,t_2]$,
and let $C_1$ and $C_2$ be the earliest collects of $p_1$ and $p_2$
that overlap $(t_1,t_2]$.
Assume that $C_1$ starts before $C_2$, and
consider the first step $\pi$ of $C_1$ in $(t_1,t_2]$.  There are
three
cases, depending on when $\pi$ occurs relative to $C_2$:
\begin{enumerate}
\item If $\pi$ occurs before $C_2$, then either it is the last step
of $C_1$ that occurs before $C_2$, or there is some later step
$\pi'$ that is the last step that occurs before $C_2$.  In either
case, $C_1$ takes a step that is extraneous for $p_2$, and we have
$m_{p_2p_1} \ge 1$.
\item If $\pi$ occurs during $C_2$, it is useful for $p_2$, 
and we again have $m_{p_2p_1} \ge 1$.
\item If $\pi$ occurs after the end of $C_2$, it either occurs
outside of any collect,
in which case it is the first operation after the end of some collect by
$p_2$ and is extraneous for
$p_2$, or it occurs during some later collect $C'_2$.  In the latter
case,
$\pi$ is useful for $p_2$,
since $C'_2$
starts after $C_2$, and thus after $C_1$.
Whether $\pi$ is extraneous for $p_2$ or useful for $p_2$, we still
have $m_{p_2p_1} \ge 1$.
\end{enumerate}

We have just shown that $m_{p_2p_1} \ge 1$ when $C_1$ starts before
$C_2$.  In the symmetric case where $C_2$ starts before $C_1$, a
symmetric argument shows that $m_{p_1p_2} \ge 1$.  Since one of these
two cases holds, we get $m_{p_1p_2} + m_{p_2p_1} \ge 1$ as claimed.

Since $m$ processes carry out at least one step in $(t_1,t_2]$, there
are ${m \choose 2}$ distinct pairs of processes $p_1,p_2$ that each
carry out at least one step in $(t_1,t_2]$.  We have just shown that
$m_{p_1p_2}+m_{p_2p_1} \ge 1$ for each such pair, and so, continuing
from (\ref{eq-lemma-progress-rises}),
\begin{eqnarray*}
\sum_p M_p(t_2) - \sum_p M_p(t_1)
&\ge& \sum_{p_1,p_2 \in S, p_1 \ne p_2} (m_{p_1p_2} + m_{p_2p_1})
\\
&\ge& \sum_{p_1,p_2 \in S, p_1 \ne p_2} 1
\\
&=& {m \choose 2}.
\end{eqnarray*}
\end{proof}

Turning to the champion, we can trivially bound the number of collects
completed during an interval of length $n-1$ by the number of active
processes:

\begin{lemma}
\label{lemma-champion-loses}
Fix some correct collect algorithm $A^*$.
Let $t_2 = t_1 + n-1$ and suppose that there are exactly $m$ processes that
carry out at least one
step in $(t_1,t_2]$.  Then $A^*$ completes at most $m$ collects during
$(t_1,t_2]$.
\end{lemma}
\begin{proof}
Since a process must carry out at least one step to complete a
collect, the only way $A^*$ can complete more than $m$ collects during
the interval is if some process $p$ completes more than one collect.
We will show that if this happens, there is an execution that
demonstrates that $A^*$ is \emph{not} correct.  It follows that if
$A^*$ is correct, then $A^*$ completes at most $m$ collects during the
interval.

Suppose that there is such a process $p$, and let $t_p > t_1$ be the time at
which $p$ first completes a collect during $(t_1,t_2]$.
Then $(t_p,t_2]$ consists of at most $n-2$ steps, and since at most
one register can be read during any one step, by the Pigeonhole
Principle there are exist (at least) two processes $q_1,q_2$ 
with the property that no
process reads a register owned by $q_1$ or $q_2$
process during $(t_p,t_2]$.  At least one of these two processes is
not $p$; call this process $q$.

Let $v$ be the value that $p$ returns for $q$'s register from its
second collect during the interval.  We will now construct a modified
execution in which $v$ is replaced by a different value $v'$ before
this collect starts.
Let $\xi$ be
the execution of $A^*$ through time $t_2$, and split $\xi$ as $\xi =
\alpha\beta$ where $\alpha$ is the prefix whose last step occurs at
time $t_p$.  Because no process reads any register owned by $q$ in
$\beta$, we can remove all steps of $q$ in $\beta$ without affecting
the execution of the other processes; let $\beta'$ be the result of
this removal.  Now construct an execution fragment $\gamma$, extending
$\alpha$, in which $q$ runs in isolation until it completes its
current collect, and then writes a new value $v'\ne v$ to its register.
Because no process reads any register owned by $q$ in $\beta'$, the
new execution $\xi' = \alpha\gamma\beta'$ is indistinguishable from
$\xi$ by any process other than $q$; in particular, $p$ still returns
$v$ for $q$ in its second collect, which starts after $q$ writes $v'$
in $\gamma$.  Thus there is an execution in which $A^*$ returns an
incorrect value, and $A^*$ is not a correct collect algorithm.
\end{proof}

Combining Lemmas~\ref{lemma-progress-rises} and
\ref{lemma-champion-loses} gives:

\begin{lemma}
\label{lemma-interval-ratio}
Let $A$ be a collect algorithm for which $\CL(A)$ and $\PL(A)$ are
bounded, and let $A^*$ be any collect algorithm.  Let $t_2 = t_1+n-1$ and
suppose $m$ processes are active in $(t_1,t_2]$.
Let $F_p$ be the progress measure for $A$ as defined in 
(\ref{eq-progress}) in Lemma~\ref{lemma-progress-implies-collects}.
Let $C$ be the number of collects completed by $A^*$ during
$(t_1,t_2]$.
Then
\begin{eqnarray}
\frac{\sum_p (F_p(t_2) - F_p(t_1)}{C}
&\ge& 
\sqrt{
\frac{n-1}{2 \PL(A) \cdot \left(\CL(A) + 2(n-1)\right)}
}
\nonumber\\
& & -
\frac{1}{4\left(\CL(A) + 2(n-1)\right)}.
\label{eq-interval-ratio}
\end{eqnarray}
\end{lemma}
\begin{proof}
\begin{eqnarray}
\frac{\sum_p (F_p(t_2) - F_p(t_1)}{C}
&\ge&
\frac{
\frac{1}{2}
\left(
    \frac{\sum_p \left(M_p(t_2) - M_p(t_1)\right)}{\CL(A) + 2(n-1)}
    +
    \frac{\sum_p \left(N_p(t_2) - N_p(t_1)\right)}{\PL(A)}
\right)
}{m} \nonumber\\
&\ge&
\frac{1}{2m}
\left(
    \frac{{m \choose 2}}{\CL(A) + 2(n-1)}
    +
    \frac{n-1}{\PL(A)}
\right)
\nonumber\\
&=&
\frac{m-1}{4\left(\CL(A) + 2(n-1)\right)}
+
\frac{1}{m}\cdot\frac{n-1}{2 \PL(A)}.
\label{eq-interval-ratio-from-m}
\end{eqnarray}

This last quantity (\ref{eq-interval-ratio-from-m}), treated as a
function of $m$, is of the form $\frac{m-1}{a} + \frac{b}{m}$, where
$a$ and $b$ are positive constants.  Thus its second derivative is
$\frac{2b}{m^3}$, which is positive for positive $m$.
It follows that (\ref{eq-interval-ratio-from-m}) is strictly convex when $m$ is
greater than $0$, and thus that it has a unique local minimum (and no
local maxima) in the
range $m \ge 0$.  This local minimum is not at $m=0$, as the second
term diverges.  So it must be at some $m > 0$ at which the first
derivative vanishes.

Taking the first derivative with respect to $m$ and setting the result to
$0$
shows that the unique point at which the first derivative vanishes is
when
\begin{displaymath}
\label{eq-min-m}
\frac{1}{4\left(\CL(A) + 2(n-1)\right)}
=
\frac{1}{m^2} \cdot \frac{n-1}{2 \PL(A)}.
\end{displaymath}
or
\begin{equation}
\label{eq-interval-ratio-best-m}
m = \sqrt{
\frac{
    2(n-1)\cdot\left(\CL(A) + 2(n-1)\right)
}{
    \PL(A)
}}.
\end{equation}
Plugging (\ref{eq-interval-ratio-best-m}) into
(\ref{eq-interval-ratio-from-m}) and simplifying gives
the right-hand side of
(\ref{eq-interval-ratio}), which, as the minimum value of
(\ref{eq-interval-ratio-from-m}) for all $m$, is a lower bound on the
left-hand side of (\ref{eq-interval-ratio}).
\end{proof}

Equation~(\ref{eq-interval-ratio}) effectively gives us the inverse of
the competitive throughput of $A$, as we can sum over all intervals in
the schedule and use
Lemma~\ref{lemma-progress-implies-collects} to translate the lower bound on
$\sum_p F_p$ to a bound on the number of collects completed by $A$.
Asymptotically, we can simplify (\ref{eq-interval-ratio}) 
further
by noting that $\CL(A)$ is always $\Omega(n)$,
and that $\PL(A)$ is likely to
be $O(n)$ for any reasonable collect algorithm $A$.
We then get:

\begin{theorem}
\label{theorem-throughput}
Let $A$ be a collect algorithm for which $\PL(A) = O(n)$ and 
$\CL(A)$ is bounded.
Then $A$ is throughput-competitive with ratio
$O\left(\sqrt{\CL(A)}\right)$.
\end{theorem}
\begin{proof}
Fix a schedule $\sigma$ of length $t$ and a request sequence $R$.
From Lemmas~\ref{lemma-progress-implies-collects} and
\ref{lemma-interval-ratio},
we have
\begin{eqnarray*}
\done(A,\sigma,R)
&\ge&
\sum_p F_p(t) - n
\\
&\ge&
\opt(\sigma)
\cdot
\sqrt{
\frac{n-1}{2 \PL(A) \cdot \left(\CL(A) + 2(n-1)\right)}
} 
\\
& &
-
\opt(\sigma)
\cdot 
\frac{1}{4\left(\CL(A) + 2(n-1)\right)}
- n
\\
&=&
\opt(\sigma) \cdot \left(
\frac{1}{O(\sqrt{\CL(A)})}
- \frac{1}{\Omega(\CL(A))}\right)
- n
\\
&=&
\opt(\sigma) \cdot \frac{1}{O(\sqrt{\CL(A)})} - n.
\end{eqnarray*}
The last term is subsumed by the additive constant, and we are left
with just the ratio $k = O\left(\sqrt{\CL(A)}\right)$.
\end{proof}

For example, applying Theorem~\ref{theorem-throughput} to the
collect algorithm of Ajtai~\etal~\cite{AADW-94} 
gives a competitive throughput of
$O(n^{3/4} \log n)$.  Similarly,
Aspnes and Hurwood~\cite{AspnesH1998}
give a randomized algorithm whose collective latency is $O(n \log^3
n)$, and use an extended version of
Theorem~\ref{theorem-throughput} to show that it has competitive
throughput $O(n^{1/2} \log^{3/2} n)$.

\subsection{Lower bound}
\label{section-lower-bound}

It is a trivial observation that any cooperative collect algorithm has
a collective latency of at least $\Omega(n)$, for the simple reason
that completing even a single collect operation requires reading all
$n$ registers.
It follows that Theorem~\ref{theorem-throughput} cannot give an upper
bound on competitive throughput better than $O(\sqrt{n})$.
This turns out to be an absolute lower bound on the competitive
throughput of any deterministic collect algorithm, as shown in
Theorem~\ref{theorem-lower-bound}, below.

\begin{theorem}
\label{theorem-lower-bound}
No deterministic algorithm for collect or write-collect
has a throughput competitiveness less
than $\Omega(\sqrt{n})$.
\end{theorem}
\begin{proof}
Fix some deterministic algorithm $A$.
We will construct a schedule in which $A$ completes $O(\sqrt{n})$
collects, while an optimal algorithm completes $\Omega(n)$.  By iterating
this construction, we get an arbitrarily long schedule in which the
ratio of collects completed by $A$ to those completed by an optimal
algorithm is $1/\Omega(\sqrt{n})$.  Since an arbitrarily long schedule
eventually overshadows any additive constant, it follows that the
competitive throughput of $A$ is at least $\Omega(\sqrt{n})$.

Choose a set $S = \{p_1, p_2, \ldots, p_m\}$ of
$m = o(n)$ processes and construct a schedule $\sigma$ in which these processes
(and no others) take steps in round-robin order.  During the first
$n-m-1$ steps of this schedule, at most $n-m-1$ registers are read,
so in particular there is some process $p \notin S$ such that no register
belonging to $p$ is read in the first $n-m-1$ steps of the execution
of $A$.

Extend $\sigma$ to a new schedule $\sigma'$ by splitting $\sigma$ into
segments where each processes takes two steps, and inserting $m+n+1$
steps by $p$ in between the first and second round of steps in each
segment.  The result looks like this:

\begin{displaymath}
\underbrace{
p_1 p_2 \ldots p_m \overbrace{p p \ldots p}^{\times m+n+1}
p_1 p_2 \ldots p_m
}_{\times \left\lfloor\frac{n-m-1}{2m}\right\rfloor}.
\end{displaymath}

This new schedule $\sigma'$ is indistinguishable from
$\sigma$ to processes in $S$.  So in an execution of $A$ under
$\sigma'$, no process in $S$ reads any register owned by $p$, 
and so no process in $S$ completes a collect.
Turning to $p$, since $p$ can complete at most one collect for each
$n-1$ steps (the minimum time to read fresh values), 
the number of collects completed by $p$ during $\sigma'$ is at most
\begin{displaymath}
\left\lfloor
\frac{(3m+n+1)\left\lfloor\frac{n-m-1}{2m}\right\rfloor}{n-1}
\right\rfloor
= O(n/m).
\end{displaymath}

In contrast, 
a better $A^*$ might proceed as follows: during each of the
$\left\lfloor\frac{n-m-1}{2m}\right\rfloor$
segments of $\sigma'$, first $p_1$ through $p_m$ write 
out timestamps (and, in the case of write-collect, their inputs).
Process $p$ then gathers these timestamps in $m$ steps (so that in can
prove that the values it reads later are fresh).  Process $p$ uses $n$
more steps to read the $n$ registers, and writes the values of
these registers, marked with the timestamps, in its last step.  During
the last $m$ steps of the segment, $p_1$ through $p_m$ read $p$'s
registers to finish their collects.  Thus an optimal $A^*$ finishes at
least $m+1$
collects per segment, for a total of at least
$(m+1)\left\lfloor\frac{n-m-1}{2m}\right\rfloor = \Omega(n)$ 
collects during $\sigma'$.

So far we have mostly demonstrated the ``granularity problem'' that
justifies the additive constant in
Definition~\ref{definition-competitive-throughput}.
To overcome this constant, we need to iterate the construction of
$\sigma'$, after extending it further to get $A$ back to a state in
which every process is about to start a new collect.

Observe that if a process has not yet completed a collect, it cannot
do so without executing at least one operation.  Let $\rho_0$ be
the shortest schedule of the form $\sigma' p p \ldots p$ such
that in Algorithm $A$, process 
$p$ has finished a collect without starting a new collect at
the end of $\rho_0$, where $p$ is as in the definition of $\sigma'$.
Note that if $p$ has completed all of its collects in
$\sigma'$, $\rho_0$ will be equal to $\sigma'$, but in general
$\rho_0$ will add as many as $O(n)$ additional steps by $p$.
Note further that extending $\sigma'$ to $\rho_0$ adds at most one
additional completed collect for $A$.

Similarly define, for each $i$ in the range $1$ to $m$, $\rho_i$ as
the shortest schedule of the form $\rho_{i-1} p_i p_i \ldots p_i$ such
that $p_i$ has finished a collect without starting a new collect at
the end of $\rho_i$.
As before, each such extension adds at most one additional completed
collect for $A$, so that the total number of collects completed by $A$
in $\rho_m$ is at most $1+m$ more than the number completed in
$\sigma'$, for a total of $O\left(m + \frac{n}{m}\right)$.

This quantity is minimized when $m =
\Theta(\sqrt{n})$, in which case $A$ completes $O(\sqrt{n})$ collects
during $\rho_m$.
Because $\rho_m$ extends $\sigma'$, the number of collects completed
by $A^*$ can only increase, so $A^*$ still completes $\Omega(n)$
collects during $\rho_m$.

Since at the end of $\rho_m$ we are in a state where every process
is about to start a collect, we may repeat the construction 
to get a sequence 
of phases, in each of which $A$ completes $O(\sqrt{n})$ collects vs.{}
$\Omega(n)$ for $A^*$.
Call the schedule consisting of $s$ such phases $\rho^s$.
Then when $n$ is sufficiently large,
$\done(A,\rho^s,R) \le s c \sqrt{n}$ for some constant $c$, while
$\done(A^*,\rho^s,R) \ge s c^* n$ for some constant $c^*$, where $R$
is a set of request sequences consisting only of collect operations.

From Definition~\ref{definition-competitive-throughput}, $A$ is
$k$-throughput-competitive only if there exists a constant $c'$ such
that for all $\rho^s$,
\begin{displaymath}
\done(A,\rho^s,R) + c'
 \ge {1 \over k} \opt_{T}(\rho^s) \ge \frac{1}{k} \done(A^*,\rho^s,R).
\end{displaymath}
Applying our previous bounds on $\done(A,\rho^s,R)$ and
$\done(A^*,\rho^s,R)$, we get
\begin{displaymath}
s c \sqrt{n} + c' \ge \frac{1}{k} s c^* n,
\end{displaymath}
and thus
\begin{displaymath}
k \ge \frac{c^* s n}{c s \sqrt{n} + c'}.
\end{displaymath}
Since this last inequality holds for all $s$, taking the limit as $s$
goes to infinity gives
\begin{displaymath}
k \ge \frac{c^*}{c} \sqrt{n} = \Omega(\sqrt{n}).
\end{displaymath}

\end{proof}

Though we concentrate on deterministic algorithms in this paper, it is
worth noting that a similar construction gives the same lower bound
for randomized algorithms with an adaptive adversary.  The main
difference is that instead of choosing $p$ to be the last process
whose register is read, we choose $p$ to have the highest expected
time at which its register is first read, and cut off a segment when
$p$'s register is in fact read.

\section{Applications}
\label{section-applications}

Armed with a throughput-competitive write-collect algorithm and
Theorem \ref{theorem-composition}, it is not hard to obtain
throughput-competitive versions of many well-known shared-memory
algorithms.  Examples include snapshot
algorithms~\cite{G6,And,AHSnap,AHR,AR},
the 
bounded round numbers
abstraction~\cite{DHW}, concurrent timestamping
systems~\cite{DolSh,DHPW,DW,GLS,Haldar93,IsPin}, and time-lapse
snapshot~\cite{DHPW}.  Here we elaborate on some simple examples.

\subsection{Atomic snapshots}
\label{section-snapshot}

For our purposes,
a {\em
snapshot} object simulates an array of $n$ single-writer registers
that support a {\em scan-update} operation, which writes a value to one
of the registers (an ``update'') and returns a vector of values for
all of the registers (a ``scan'').
A scan-update is distinguished from
the weaker write-collect operation of
Section~\ref{section-write-collect}
by a much stronger
serialization condition; informally, this says that the vector of
scanned values must appear to be a picture of the registers at some
particular instant during the execution.
As with write-collect,
we are combining what in some
implementations may be a separate cheap operation (the update) with an
expensive operation (the scan).\footnote{A similar combined operation
appears, with its name further abbreviated to \buzz{scate}, in~\cite{AR}.}

Snapshot objects are very useful tools for constructing
more complicated shared-memory algorithms, and they have been
extensively studied~\cite{G6,And,AHSnap,AHR}
culminating in the
protocol of Attiya and Rachman~\cite{AR} which uses only $O(\log n)$
alternating writes and collects to complete a scan-update operation,
giving $O(n \log n)$ total work.

\begin{sloppypar}
We will apply Theorem \ref{theorem-composition} to get a competitive
snapshot.  Let $T$ be a
snapshot object and $U$ a write-collect object.  Because a
scan-update can be used to simulate a write-collect or collect, 
we have
$\opt_T(\sigma) \leq \opt_U(\sigma)$ for any schedule
$\sigma$, and
so scan-update is feasible relative to write-collect.

Now let $A$ be the Attiya-Rachman snapshot algorithm, and let $B$ be a
throughput-competitive implementation of write-collect.
Let $R$ be a set of request sequences consisting of scan-update
operations.
Since each process in the Attiya-Rachman snapshot
algorithm completes one scan-update for every $O(\log n)$
write-collects, we have
$\done(B,\sigma,R_A) 
\le O(\log n)\cdot \done(A\circ B,\sigma,R) 
+ O(n \log n)$,
where the additive term accounts for write-collect operations that are
part of scan-updates that have not yet finished at the end of
$\sigma$.
So we have:

\begin{displaymath}
{\done(A \circ B,\sigma,R) + O(n) \over \done(B,\sigma,R_A)}
\geq {1 \over O(\log n)}
\geq {1 \over O(\log n)} \cdot
 \frac{\opt_T(\sigma)}{\opt_U(\sigma)},
\end{displaymath}
since $frac{\opt_T(\sigma)}{\opt_U(\sigma)} \le 1$.
Applying Definition~\ref{definition-relative-competitiveness},
the Attiya-Rachman snapshot is 
$O(\log n)$-throughput-competitive relative to write-collect.  
By Theorem
\ref{theorem-composition}, plugging in any
$k$-throughput-competitive implementation of write-collect gives
an $O(k \log n)$-throughput-competitive snapshot protocol.
For example, if we use the $O(n^{3/4} \log n)$-competitive protocol of
Section~\ref{section-implementation-of-write-collect}, we get an
$O(n^{3/4} \log^2 n)$-competitive snapshot.
\end{sloppypar}

\subsection{Bounded round numbers}
\label{section-round-numbers}

A large class of wait-free algorithms that communicate via
single-writer multi-reader atomic registers have a
communication structure
based on {\em asynchronous rounds}.
Starting from round 1, at each round, the process performs a
computation, and then advances its round number and proceeds to
the next round.  A process's actions do not depend on its exact round
number, but only on the distance of its current round number from
those of other processes.  Moreover, the process's actions are not
affected by any process whose round number lags behind its own by more
than a finite limit. The round numbers increase unboundedly over the
lifetime of the system. 

Dwork, Herlihy and Waarts~\cite{DHW}
introduced the \buzz{bounded round numbers}
abstraction, which can be plugged into any algorithm that uses round
numbers in this fashion, transforming it into a bounded
algorithm.  The bounded round numbers implementation in
\cite{DHW} provides four operations of varying difficulty; however,
the use of these operations is restricted.  As a result, we
can coalesce these operations into a single operation, an
\buzz{advance-collect}, which advances the current process's round
number to the next round and collects the round numbers of the other
processes.  
Using their implementation, only $O(1)$ alternating writes
and collects are needed to implement an advance-collect.

Again we can apply Theorem \ref{theorem-composition}.  Let $T$ be a
an object providing the
advance-collect operation, and let $U$ be a write-collect object.  
Because an
advance-collect must gather information from every process in the
system, it implicitly contains a collect, and 
$\opt_T(\sigma) \leq \opt_U(\sigma)$ for all schedules $\sigma$.  An
argument similar to that used above for the Attiya-Rachman snapshot
thus shows that plugging a $k$-throughput-competitive implementation of
write-collect into the Dwork-Herlihy-Waarts bounded round numbers
algorithm gives an $O(k)$-throughput-competitive algorithm.
Using the write-collect algorithm of
Section~\ref{section-implementation-of-write-collect} thus gives an
$O(n^{3/4} \log n)$-competitive algorithm.


%

\section{Conclusions}
\label{section-conclusions}

\begin{sloppypar}
We have given a new measure for the competitive performance of
distributed algorithms, which improves on the competitive latency
measure of Ajtai~\etal~\cite{AADW-94} by allowing such algorithms to
be constructed compositionally.  We have shown that the cooperative collect
algorithm of~\cite{AADW-94}
is $O(n^{3/4} \log^{3/2} n)$-competitive by this measure, from which
we get an $O(n^{3/4} \log^{5/2} n)$-competitive atomic snapshot by
modifying
the protocol of~\cite{AR}, and an $O(n^{3/4} \log^{3/2}
n)$-competitive bounded round numbers protocol by modifying the protocol
of~\cite{DHW}.  These modifications require only replacing the
collect subroutine used in these protocols with a cooperative collect
subroutine, and the proof of competitiveness does not require
examining the actual working of the modified protocols in detail.
We believe that a similar straightforward substitution could give
competitive versions of many other distributed protocols.
\end{sloppypar}

We discuss some related approaches to analyzing the competitive ratio
of distributed algorithms in Section~\ref{section-related-work}.
Some possible extensions of the present work are mentioned in
Section~\ref{section-variations}.  

Finally, we note that 
competitive ratios of $\tilde{O}(n^{3/4})$ are not very good, but they are
not too much worse than Theorem~\ref{theorem-lower-bound}'s lower bound of
$\Omega(n^{1/2})$.  We describe some related work that gets closer to this
bound (and, for a modified version of the problem, breaks it) in
Section~\ref{section-improvements}.

\subsection{Related work}
\label{section-related-work}

A notion related to allowing only other distributed algorithms as
champions is the very nice idea of comparing algorithms with partial
information only against other algorithms with partial information. This
was introduced by Papadimitriou and Yannakakis~\cite{PY-93} in the
context of linear programming; their model corresponds to a distributed
system with no communication.  A generalization of this approach has
recently been described by Koutsoupias and Papadimitriou~\cite{KP-94}.

In addition, there is a long history of interest in {\em optimality}
of a distributed algorithm given certain conditions, such as a
particular pattern of failures~\cite{DRS,DM,HMW,MT,Neiger90,NT}, or a
particular pattern of message delivery~\cite{AHerzR,FM,PR}.  
In a sense, work on optimality envisions a fundamentally different role
for the adversary in which it is trying to produce bad performance 
for both
the candidate and champion
algorithms; 
in contrast, the adversary used in competitive analysis
usually cooperates with the champion.

Nothing in the literature corresponds in generality
to our notion of relative competitiveness 
(Definition \ref{definition-relative-competitiveness})
and the composition theorem (Theorem
\ref{theorem-composition}) that uses it. Some examples of
elegant specialized constructions of competitive algorithms from other
competitive algorithms in a distributed setting are the {\em
natural potential function} construction of 
Bartal \etal~\cite{BFR-92} 
and the distributed paging algorithm of 
Awerbuch \etal~\cite{AwerbuchBF1998}. 
However, not only do these constructions
depend very much on the particular details of the problems being solved,
but, in addition, they permit no concurrency, {\em i.e.} they assume that no
two operations are ever in progress at the same time. (This assumption does
not hold in general in typical distributed systems.) In contrast, the
present work both introduces a general construction of compositional competitive
distributed algorithms {\em and} does so in the natural distributed setting
that permits concurrency. 

\subsection{Variations on competitiveness}
\label{section-variations}

Our work defines compositional competitiveness and relative competitiveness
by distinguishing between two sources of nondeterminism, one of
which is shared between the on-line and off-line algorithms, 
{\em i.e.} the schedule, and the other is not, {\em i.e.} the input. One
can define analogous notions to compositional competitiveness and to relative
competitiveness by considering \emph{any} two sources of nondeterminism, one of
which is shared between the on-line and off-line algorithms, and one
that is not. 
This leads to a general notion of \buzz{semicompetitive analysis},
which has been described in
a survey
paper by the first author~\cite{Dagstuhl},
based in part on the present work.

\subsection{Improved collect algorithms}
\label{section-improvements}

Since the appearance of the conference version of this paper,
Aspnes and Hurwood \cite{AspnesH1998} and Aumann \cite{Aumann} have
shown that weakening some of the requirements of the model used here
can greatly improve performance.  

In particular, Aspnes and Hurwood~\cite{AspnesH1998}
have shown that with an adversary whose knowledge of the system state
is limited, collects can be performed with a near-optimal $O(n^{1/2}
\log^{3/2} n)$ competitive ratio in the throughput-competitiveness
model.  Aumann~\cite{Aumann} has
shown that, for some applications, the freshness requirement can be
weakened to allow a process to obtain a value that is out-of-date for
its own collect, but that was current at the start of some other
process's collect.  He shows that with this weakened requirement an
algorithm based on the Aspnes-Hurwood algorithm can perform collects
with a competitive ratio of only $O(\log^3 n)$.

\section{Acknowledgments}
\label{section-acknowledgments}

We are indebted to Miki Ajtai and Cynthia Dwork for very helpful
discussions, and to Maurice Herlihy on helpful comments on the
presentation of this work. We also thank Amos Fiat for his
encouragement, and the anonymous referees for very detailed and
helpful comments on an earlier draft of this work.

\bibliographystyle{plain}
\bibliography{journal}

\end{document}